\documentclass[letterpaper]{article}
\usepackage{amsmath}
\usepackage{amssymb}
\usepackage{graphicx}

\newcommand{\ycp}{Yb$_2$Co$_{12}$P$_7$}

\begin{document}

\title{Transport, magnetic, and thermal properties of non-centrosymmetric Yb$_2$Co$_{12}$P$_7$}
\author{J.\ J.\ Hamlin, M.\ Janoschek,$^{\dag}$ R.\ E.\ Baumbach,$^{\dag}$\\ B.\ D.\ White, and M.\ B. Maple$^*$\\
\textit{Department of Physics, University of California, San Diego},\\ 
\textit{La Jolla, CA 92093}}

\maketitle

\begin{abstract}
  We report magnetization and specific heat measurements down to 2 K and electrical resistivity down to millikelvin temperatures on polycrystalline samples of the non-centrosymmetric compound Yb$_2$Co$_{12}$P$_7$.  In addition to the previously reported ferromagnetic ordering of the cobalt sub-lattice at $T_C = 136$ K we find a magnetic transition below $T_M = 5$ K that is likely associated with ordering of the Yb ions.  The broad nature of the specific heat anomaly suggests disordered magnetism and possible short range correlations well above $T_M$.
\end{abstract}

\section{Introduction}
\let\thefootnote\relax\footnotetext{$^{\dag}$current address: Los Alamos National Laboratory, Los Alamos, NM 87545, USA}
\let\thefootnote\relax\footnotetext{$^*$Corresponding author. email: mbmaple@physics.ucsd.edu}
The hexagonal, non-centrosymmetric, Zr$_2$Fe$_{12}$P$_7$-type crystal structure (space group P$\overline{6}$)~\cite{ganglberger_1968} is exhibited by a large number of compounds with the chemical formula $M_2T_{12}Pn_7$ ($M$ = Li, Na, Ca, Mg, Ti-Hf, Nb, Sc, Y, La-Lu; $T$ = Mn, Fe, Co, Ni, Ru; and $Pn$ = P, As)~\cite{jeitschko_1978_1,jeitschko_1993,hellmann_2001}.  Signatures of strong electronic correlations have been found in certain members of this family of compounds for which the $M$ ion is a rare earth element with an unstable valence.  Electrical resistivity, magnetic susceptibility, and specific heat measurements on Yb$_2$Ni$_{12}$(P,As)$_7$ show evidence of an Yb valence between 2.2 and 2.8 and a moderately enhanced electronic contribution to the specific heat, $\gamma$~\cite{cho_1998_1}. X-ray absorption measurements on Ce$_2$Ni$_{12}$P$_7$ indicate a cerium valence of 3.2~\cite{babizhetskyy_2007_1}.  Yb$_2$Fe$_{12}$P$_7$ exhibits an enhanced $\gamma$ and, under the application of magnetic field $H$, exhibits a crossover from a magnetically ordered non-Fermi-liquid (NFL) phase at low $H$ to another NFL phase at higher $H$~\cite{baumbach_2010_1}.  Sm$_2$Fe$_{12}$P$_7$ appears to be a rare example of a Sm-based, heavy-fermion ferromagnet~\cite{janoschek_2011_1}.

One of the notable characteristics of this class of compounds is that the transition metal sub-lattice can be tuned from non-magnetic ($T$ = Fe or Ni) to magnetic ($T$ = Co)~\cite{jeitschko_1978_1,reehuis_1989_1,ghadraouli_1990,raffius_1991,zeppenfeld_1993,reehuis_1998}.  The $T$ = Co compounds exhibit ferromagnetic ordering of the cobalt ions where the Curie temperature $T_C$ is usually $\sim 150$ K~\cite{reehuis_1989_1}.  Additional magnetic transitions are found at lower temperatures for the cobalt-phosphide compounds that incorporate a trivalent, magnetic, rare-earth element as the $M$ ion.  A detailed magnetization and neutron diffraction study of The $M_2$Co$_{12}$P$_7$ compounds with $M$ = Pr, Nd, and Ho showed that the lower temperature transitions are due to ferromagnetic ordering of the rare earth ions~\cite{reehuis_1997_1}.  The Co moments align parallel to the crystal hexagonal axis near 150 K, while the rare earth moments for the Pr, Nd, and Ho compounds align nearly perpendicular, anti-parallel, and parallel to the Co moments, respectively.  Given the interesting behavior observed in the other Yb-based members of this family of compounds, Yb$_2$Ni$_{12}$(P, As)$_7$~\cite{cho_1998_1} and Yb$_2$Fe$_{12}$P$_7$~\cite{baumbach_2010_1}, we undertook a study of the low temperature properties of \ycp.

\section{Experimental Details}
Polycrystalline samples of \ycp\ were prepared from elemental Yb (small dendritic pieces), Co (powder), and P (small lumps).  The starting materials were sealed under vacuum in quartz tubes, slowly heated to 1000 $^{\circ}$C, and held at this temperature for 3 days.  After the initial reaction the sample was ground to a powder, pressed into a pellet, and fired at 1135 $^{\circ}$C for 3 days.  Finally, the sample was again re-ground and fired at 1135 $^{\circ}$C for 3 days.

Powder x-ray diffraction measurements were performed using a Bruker D8 Discover system utilizing CuK$_{\alpha}$ radiation.  Magnetization and specific heat measurements were performed from room temperature down to $\sim 2$ K in Quantum Design MPMS and PPMS systems, respectively.  For electrical resistivity measurements, the samples were cut into bars and four gold leads were attached using Epotek silver epoxy.  The resistivity was measured down to $\sim 1$ K in a home-built $^4$He system, while temperatures down to $\sim 50$ mK were achieved in an Oxford Instruments Kelvinox $^3$He-$^4$He dilution refrigerator.  At the lowest temperatures, the electrical resistivity measurements were repeated using several different excitation currents in order to ensure that there was no spurious heating of the sample.

Analysis of the x-ray diffraction measurements confirms the Zr$_2$Fe$_{12}$P$_7$-type crystal structure with lattice constants $a = 9.023$ \AA{} and $c = 3.580$ \AA{}, nearly identical to those reported earlier~\cite{reehuis_1989_1}.  Initial x-ray results pointed to contamination by a single impurity phase of a few percent Yb$_2$O$_3$.  However, we found that the Yb$_2$O$_3$ phase developed when, in preparation for the powder diffraction measurements, the sample was ground in air.  By performing the grinding and x-ray measurements under inert atmosphere, the Yb$_2$O$_3$ phase could be greatly reduced, indicating that the as-grown samples are nearly phase pure.
\begin{figure}
  \begin{center}
    \includegraphics[width=0.9\columnwidth]{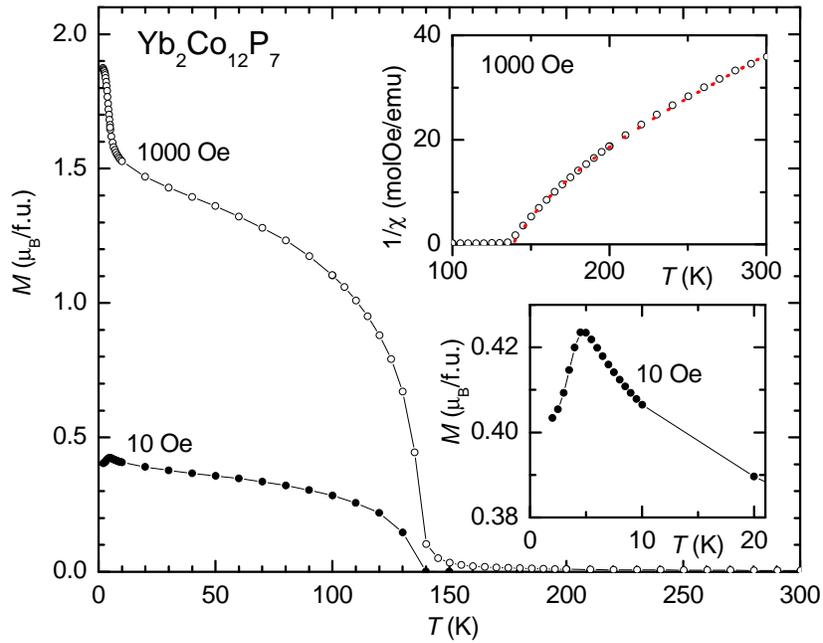}
  \end{center}
  \caption{Magnetization versus temperature data measured during field cooling under applied fields of 10 Oe and 1000 Oe.  (upper inset) Inverse susceptibility versus temperature.  The dashed red line is a fit to the data with the sum of two Curie-Weiss laws as described in the text. (lower inset) Low temperature region of the 10 Oe magnetization showing a second magnetic transition $T_M = 5$ K.}
  \label{fig:fig1}
\end{figure}

\section{Results and Discussion}
Figure~\ref{fig:fig1} presents the results of field-cooled magnetization versus temperature measurements.  The onset of ferromagnetic order is clearly visible near $T_C = 140$ K.  An additional magnetic transition appears near $T_M = 5$ K.  In an applied field of 10 Oe, this transition is manifested as a drop in the susceptibility (see inset of Figure~\ref{fig:fig1}), suggesting antiferromagnetic ordering.  However, when the measurement is performed in 1000 Oe, the susceptibility instead increases below $T_M$, more consistent with field-induced ferromagnetic ordering.  The differences between the 10 Oe and 1000 Oe susceptibility curves raise the possibility of a metamagnetic transition under applied field.

An analysis of the inverse susceptibility in terms of a Curie-Weiss law is complicated by the fact that \ycp\ has two species of magnetic ions with different moments and different ordering temperatures.  The susceptibility can then be modeled as
\begin{equation}
  \chi(T)=\dfrac{C_{Yb}}{T-\Theta _{Yb}}+\dfrac{C_{Co}}{T-\Theta _{Co}},
\end{equation}
where the Curie-Weiss temperatures are $\Theta _{Yb}$ and $\Theta _{Co}$ and the effective moments are given by $\mu_{eff}^{Yb} = 2.82\sqrt{C_{Yb}/2}$ and $\mu_{eff}^{Co} = 2.82\sqrt{C_{Co}/12}$.  Allowing all four parameters to vary while fitting the data above 140 K results in unrealistic values of the effective moments.  Instead, we fit the Curie-Weiss temperatures while fixing $\mu_{eff}^{Yb} = 4.53$ $\mu _B$ at the Hund's rule value for Yb$^{3+}$ and $\mu_{eff}^{Co} = 1.14$ $\mu _B$ at the value estimated from earlier studies of Lu$_2$Co$_{12}$P$_7$~\cite{reehuis_1989_1}.  This yields $\Theta _{Yb} = -32.9$ K and $\Theta _{Co} = 138.6$ K and the resulting curve is plotted in the upper inset to Figure~\ref{fig:fig1}.  While the exact value of $\Theta _{Yb}$ depends somewhat on the value selected for the corresponding effective moment, any reasonable estimate of $\mu_{eff}^{Yb}$ yields a negative Curie-Weiss temperature that is sizable compared to the observed ordering temperature $T_M = 5$ K, suggesting the possibility of frustrated magnetism.  Alternatively, crystal field effects, valence fluctuations, or Kondo physics could be responsible for this behavior.

\begin{figure}
  \begin{center}
    \includegraphics[width=0.9\columnwidth]{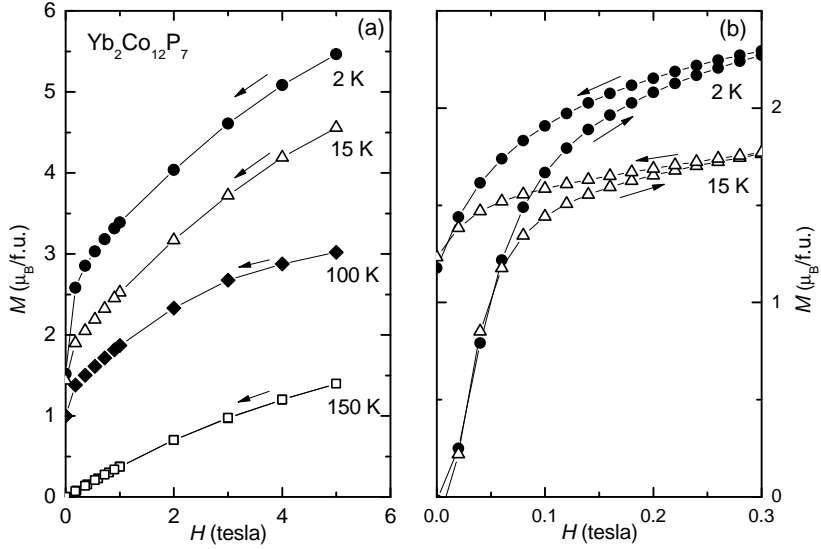}
  \end{center}
  \caption{(a) Magnetization versus field measured upon reducing the field from 5 tesla. (b) Comparison of the low field region of the magnetization versus temperature above and below $T_M$.}
  \label{fig:fig2}
\end{figure}

Figure~\ref{fig:fig2} shows the results of isothermal magnetization $M$ versus magnetic field $H$ scans.  The curves below $T_C$ can be broken into two components, one of which saturates in relatively small fields of 1000 Oe and another part which remains unsaturated by 5 tesla.  This may be attributed to the fact that the applied field is not parallel to the easy axis for every crystallite in the polycrystalline sample~\cite{holstein_1941}, precluding a straightforward determination of the exact magnitude of the ordered moments from the present magnetization data.  However, by noting that the 2 K and 15 K magnetization curves differ only by a constant at higher fields, it is clear that the magnetic component due to the transition at $T_M = 5$ K saturates by a few thousand Oe.  Moreover, the field dependence of this component is reversible since the residual magnetization upon reducing the field to zero above or below $T_M$ are nearly identical (see Figure~\ref{fig:fig2}b).

\begin{figure}
  \begin{center}
    \includegraphics[width=0.9\columnwidth]{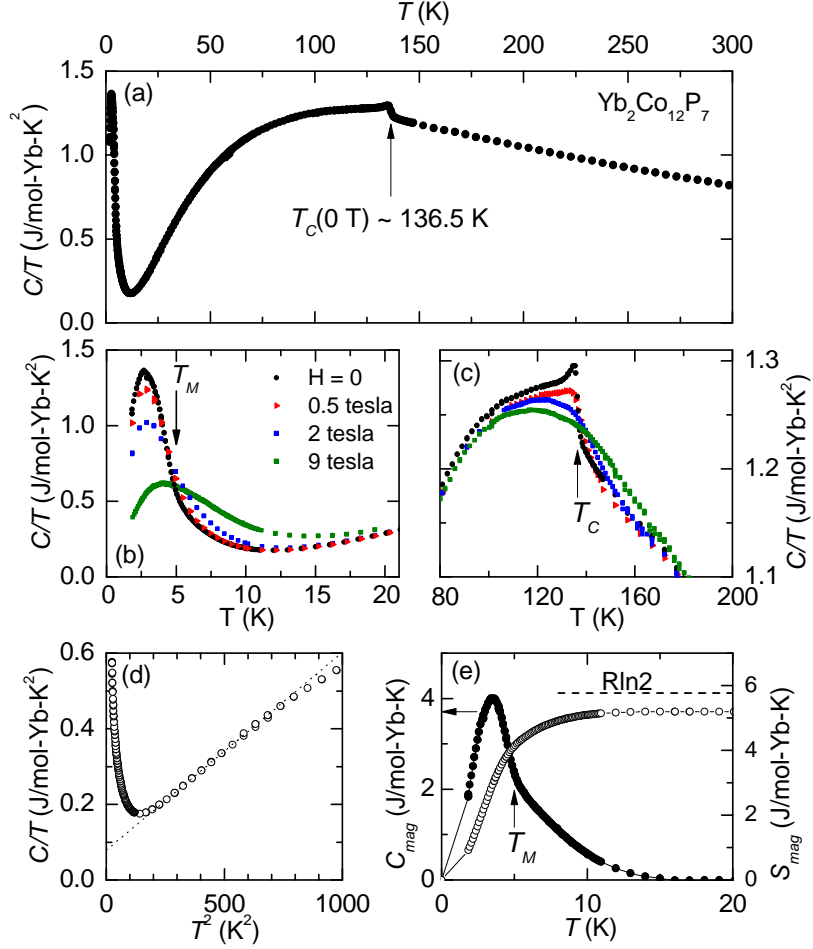}
  \end{center}
  \caption{(a) Specific heat $C$ divided by temperature $T$ versus $T$ measured in magnetic fields from 0-9 tesla. (b) and (c) highlight the behavior in the vicinity of $T_M$ and $T_C$, respectively.  Both transitions broaden under the application of magnetic field, consistent with ferromagnetic order.  (d) Low temperature region plotted versus $T^2$.  The linear fit provides an estimate of the electronic specific heat coefficient $\gamma$ and phonon contribution $\beta$ as described in the text.  (e) Magnetic contribution to the specific heat $C_{mag}$ (filled circles, left axis) and entropy $S_{mag}$ (open circles, right axis) at low temperature.}
  \label{fig:fig3}
\end{figure}

Specific heat $C$ divided by temperature $T$ data in magnetic fields from 0-9 tesla are plotted in Figure~\ref{fig:fig3}.  An anomaly due to ferromagnetic ordering of the cobalt ions is visible at $T_C = 136.5$ (Figure~\ref{fig:fig3}c).  Application of a magnetic field smears out this transition, as may be expected for a ferromagnetic transition.  At lower temperatures, a second anomaly appears as a broad maximum (Figure~\ref{fig:fig3}b).  In zero magnetic field, the ordering temperature as determined by the inflection point in the $C/T$ curve is $T^C_M = 5$ K, in good agreement with that determined from magnetic susceptibility.  The low temperature anomaly broadens and shifts to higher temperatures with the application of magnetic field, again suggestive of ferromagnetism.  These measurements also rule out any significant contamination by Yb$_2$O$_3$ which is known to exhibit a large, sharp specific heat anomaly near 2.5 K~\cite{Li_1994}.  No sign of such a transition is apparent in the present data.

The specific heat well below the Debye temperature $\Theta _D$ can be described by $C(T) = C_{mag}(T) + \gamma T + \beta T^3$, where $C_{mag}(T)$ is the magnetic contribution, $\gamma$ is electronic specific heat coefficient and $\beta$ gives the lattice contribution from the lowest order term of the Debye function.  The values of $\beta$ and $\gamma$ can be obtained from the slope and intercept, respectively of a linear fit to a plot $C/T$ versus $T^2$ at low temperatures (but above the magnetic anomaly).  This is shown in Figure~\ref{fig:fig3}d and results in $\beta = 0.53$ mJ/mol-Yb-K$^4$ and a moderately enhanced electronic contribution $\gamma = 77$ mJ/mol-Yb-K$^2$.  Specific heat measurements on Hf$_2$Co$_{12}$P$_7$ indicate an electronic contribution that may be as large as $\gamma \sim 50$ mJ/mol-Hf-K$^2$~\cite{in_prep}.  This suggests that a substantial portion of $\gamma$ in \ycp\ may derive from a large density of states at the Fermi level associated with the Co $d$-electrons rather than from hybridized Yb $f$-electron states.

An estimate of of the Debye temperature $\Theta_D = (r \times 1944/\beta)^{1/3}$, where $r = 21$ is the number of atoms per formula unit, yields $\Theta _D = 425$ K.  Approximating the phonon contribution as $\beta T^3$ should therefore be reasonable below $\sim 40$ K (10\% of $\Theta _D$). By subtracting the lattice ($\beta T^3$) and electronic ($\gamma T$) contributions to the specific heat we arrive at an estimate of $C_{mag}(T)$, which is plotted in Figure~\ref{fig:fig3}e.  An increase in $C_{mag}(T)$ begins already below $\sim 15$ K, nearly three times higher that the ordering temperature as determined from magnetization and electrical resistivity.  This may indicate that short range magnetic correlations develop well above $T_M$.  The line connecting to the origin indicates a rough extrapolation of $C_{mag}(T)$ to lower temperatures for the purpose of estimating the entropy.  Integrating $C_{mag}(T)/T$ yields the magnetic entropy (Figure~\ref{fig:fig3}e), which saturates at 90\% of R$\ln 2$ per Yb ion by $\sim 10$ K.  This is consistent with the magnetic transition at $T_M$ involving ordering of Yb$^{3+}$ ions in a Kramers doublet ground state.

\begin{figure}
  \begin{center}
    \includegraphics[width=0.9\columnwidth]{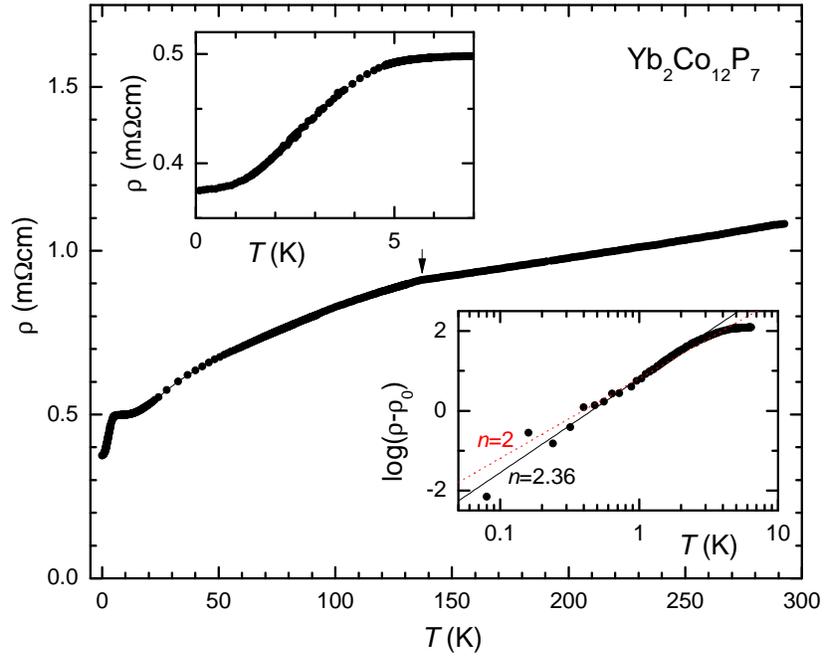}
  \end{center}
  \caption{Electrical resistivity versus temperature for polycrystalline \ycp.  The arrow indicates the ferromagnetic ordering temperature $T_C$.  (upper inset) Resistivity at low temperature.  (lower inset) Power law, $\rho(T) = \rho_0 + AT^n$, dependence of the low temperature electrical resistivity.  The best fit is obtained with $n=2.36$ (black line) although the data can also be reasonably well fit with $n=2$ (dashed red line).}
  \label{fig:fig4}
\end{figure}

Figure~\ref{fig:fig4} shows the electrical resistivity versus temperature.  At the highest temperatures the electrical resistivity is roughly linear, as is typical for a normal metal.  A kink appears in the electrical resistivity at 135.6 K, nearly the same temperature identified as the Curie temperature from the the magnetization and specific heat measurements.  Below the Curie temperature the slope increases significantly and displays negative curvature down to $\sim 25$ K.  This negative curvature may be due to either $s$-$d$ scattering as described by Mott and Jones~\cite{mott_jones_1958} or crystalline electric field (CEF) effects. The latter possibility is particularly likely given the entropy, which is consistent with CEF effects splitting the Hund's rule multiplet to yield a doublet ground state.  Another possibility is that the negative curvature results from the onset of coherent scattering processes from hybridized Yb ions.  Based on this assumption, the shoulder in the resistivity at $\sim 40$ K would provide a rough estimate of the Kondo coherence temperature.  A similar shoulder was also found at $\sim 30$ K in Yb$_2$Fe$_{12}$P$_7$~\cite{baumbach_2010_1}.

The upper left inset of Figure~\ref{fig:fig4} highlights the drop in the resistivity that occurs below $T_M = 5$ K.  Below the transition, the electrical resistivity is consistent with power law behavior, $\rho (T) = \rho_0 + AT^n$.  The value of $n$ can be determined from the slope of a plot of $\log (\rho - \rho_0)$ versus $\log (T)$, where $\rho _0$ is selected to maximize the range of linear behavior extending from low temperature.  The best fit to the data is obtained for $n = 2.36$, with the power law behavior extending up to roughly 3 K, as illustrated in the lower inset to Figure~\ref{fig:fig4}.  The data can also be fit over a somewhat smaller temperature range with an $n = 2$ power law.  A $T^2$ temperature dependence of the electrical resistivity could be explained either by ferromagnetic magnon scattering or by Fermi liquid behavior, as is found for many heavy fermion materials.  To help distinguish between these possibilities, we consider the Kadowaki-Woods ratio $R_{KW} = A/\gamma ^2$, which gives a measure of the relationship between the coefficient of electronic specific heat $\gamma$ and the coefficient $A$ of the $T^2$ contribution to the electrical resistivity~\cite{kadowaki_1986}.  For many Yb-based heavy fermion compounds it has been found that $R_{KW} \sim 0.36 \times 10^{-6}$ $\mu \Omega$cm(mol-Yb-K/mJ)$^2$~\cite{tsujii_2005}.  For \ycp, we find $A = 7.9$ $\mu \Omega$cm/K$^2$ and $\gamma = 77$ mJ/mol-Yb-K$^2$, resulting in a value $A/\gamma ^2 = 1.3 \times 10^{-3}$ $\mu \Omega$cm(mol-Yb-K/mJ)$^2$.  Because this value is much larger than that typically found for Yb-based heavy fermion compounds, we conclude that the low temperature electrical resistivity is likely dominated by magnetic scattering rather than heavy Fermi liquid behavior.

\section*{Acknowledgements}
Synthesis and screening for superconductivity were carried out under the auspices of AFOSR-MURI, Grant FA9550-09-1-0603.  Physical properties characterization was supported by the Department of Energy (DOE), Grant DE-FG02-04-ER46105.  Dilution refrigerator measurements were supported by the National Science Foundation (NSF), Grant DMR-0802478.  M. Janoschek gratefully acknowledges financial support from the Alexander von Humboldt foundation.

\end{document}